\documentclass{article}

\usepackage{PRIMEarxiv}

\usepackage[utf8]{inputenc} % allow utf-8 input
\usepackage[T1]{fontenc}    % use 8-bit T1 fonts
\usepackage{hyperref}       % hyperlinks
\usepackage{url}            % simple URL typesetting
\usepackage{booktabs}       % professional-quality tables
\usepackage{amsfonts}       % blackboard math symbols
\usepackage{nicefrac}       % compact symbols for 1/2, etc.
\usepackage{microtype}      % microtypography
\usepackage{lipsum}
\usepackage{fancyhdr}       % header
\usepackage{graphicx}       % graphics
\graphicspath{{media/}}     % organize your images and other figures under media/ folder
\usepackage{subcaption}
\title{HelixDesign-Binder: A Scalable Production-Grade Platform for Binder Design Built on HelixFold3}

\author{
  Jie Gao, Jun Li, Jing Hu, Shanzhuo Zhang, Kunrui Zhu, Yueyang Huang, Xiaonan Zhang, Xiaomin Fang\thanks{Corresponding author. Email: fangxiaomin01@baidu.com } \\
  PaddleHelix team, Baidu Inc. \\
}

\begin{document}
\maketitle

\begin{abstract}
Protein binder design is central to therapeutics, diagnostics, and synthetic biology, yet practical deployment remains challenging due to fragmented workflows, high computational costs, and complex tool integration. We present HelixDesign-Binder, a production-grade, high-throughput platform built on HelixFold3 that automates the full binder design pipeline—from backbone generation and sequence design to structural evaluation and multi-dimensional scoring. By unifying these stages into a scalable and user-friendly system, HelixDesign-Binder enables efficient exploration of binder candidates with favorable structural, energetic, and physicochemical properties. The platform leverages Baidu AI Cloud’s high-performance infrastructure to support large-scale design and incorporates advanced scoring metrics, including ipTM, predicted binding free energy, and interface hydrophobicity. Benchmarking across six protein targets demonstrates that HelixDesign-Binder reliably produces diverse and high-quality binders, some of which match or exceed validated designs in predicted binding affinity. HelixDesign-Binder is accessible via an interactive web interface in \href{https://paddlehelix.baidu.com/app/all/helixdesign-binder/forecast}{PaddleHelix platform}, supporting both academic research and industrial applications in antibody and protein binder development.
\end{abstract}

% keywords can be removed

\section{Introduction}
Protein binder design is a foundational technique in therapeutic development, molecular diagnostics, and synthetic biology. By engineering proteins that selectively bind to target biomolecules, researchers can modulate biological pathways, facilitate targeted delivery, and construct molecular sensors. Structure-based binder design aims to improve binding specificity and stability by generating candidate sequences that adopt desirable conformations and interact favorably with a given target.

Recent breakthroughs in structure prediction,  most notably AlphaFold-Multimer \cite{evans2021protein}, AlphaFold3 \cite{abramson2024accurate}, and \href{https://paddlehelix.baidu.com/app/all/helixfold3/forecast}{HelixFold3} \cite{liu2024technical}, have substantially advanced our ability to model protein–protein complexes with high accuracy. Recent work such as AlphaProteo \cite{zambaldi2024novo} and BindCraft \cite{pacesa2024bindcraft} utilize these structure prediction models in silico screening of designed sequences for favorable binding conformations, making it feasible to incorporate structural evaluation early in the binder design process. In a typical pipeline, the 3D structure of a target protein is fixed as the design scaffold. Tools like RFDiffusion \cite{watson2023novo, ahern2025atom} are first used to generate binder backbone structures that are spatially compatible with the target surface. Then, inverse folding models such as ESM-IF \cite{hsu2022learning} and ProteinMPNN \cite{dauparas2022robust} are applied to design sequences expected to fold into these backbones and maintain favorable interactions with the target. Structural prediction tools assess whether the designed binders adopt realistic conformations and correctly engage the target interface. Downstream filters, such as clash detection, binding pose analysis, and interaction scoring with FoldX \cite{schymkowitz2005foldx} or PRODIGY \cite{xue2016prodigy}, help prioritize high-quality candidates for experimental validation.

However, deploying such workflows in practice is hindered by several factors. First, integrating diverse tools—each with its own interface, input requirements, and parameter conventions—leads to fragmented pipelines that are challenging to configure and maintain, particularly for researchers without extensive programming expertise. Second, tuning parameters across multiple modules is non-trivial due to the lack of standardized guidelines. Finally, high-accuracy structural prediction demands substantial computational resources, and limited sampling can constrain sequence diversity, reducing the likelihood of identifying high-affinity binders. Empirical evidence \cite{bennett2023improving, hsu2022learning, dauparas2022robust, hayes2025simulating} indicates that broader sequence exploration significantly enhances the probability of discovering stable, high-affinity candidates.

To address these challenges, we present HelixDesign-Binder, a scalable production-grade binder design platform built upon HelixFold3. Unlike traditional workflows where backbone generation, sequence design, and structural evaluation are typically handled by disparate modules requiring extensive manual configuration and resource coordination, HelixDesign-Binder offers a fully integrated and automated pipeline. It unifies backbone construction, structure-based sequence design, structure prediction, and multi-dimensional scoring into a cohesive framework that scales efficiently for large-scale binder design. Engineered to support both academic research and industrial deployment, the platform enables comprehensive evaluation of candidate binders across several critical dimensions, including sequence plausibility, structural quality, and physicochemical compatibility. This tight integration significantly reduces system complexity and operational overhead, while enhancing the overall efficiency, scalability, and robustness of the design process. Extensive benchmarking across a wide range of protein targets demonstrates that HelixDesign-Binder enables systematic and high-fidelity exploration of the sequence space, consistently yielding structurally plausible and functionally promising binder candidates. The platform is accessible via the \href{https://paddlehelix.baidu.com/app/all/helixdesign-binder/forecast}{PaddleHelix platform}, allowing researchers to easily initiate binder design tasks and interact with results through a user-friendly interface.

Key features of the HelixDesign-Binder are summarized as follows:
\begin{itemize}
\item \textbf{Precise Structural Evaluation via HelixFold3:}
Powered by HelixFold3, the platform delivers atomic-level predictions of protein-binder interactions with accuracy on par with AlphaFold3, and surpasses it in some cases. Incorporating prior knowledge like reference structures and interaction constraints further improves prediction precision, enabling early identification of promising binders and accelerating design.
\item \textbf{High-Throughput Design and Evaluation}:
HelixDesign-Binder leverages the computational power of Baidu AI Cloud’s high-performance computing (HPC) platform to enable rapid design and evaluation of thousands of binder candidates. This capability supports comprehensive exploration of the sequence space, thereby increasing the probability of identifying candidates with favorable binding affinities and interface qualities.
\item \textbf{Multi-Dimensional Scoring and Filtering}:
HelixDesign-Binder integrates multiple evaluation dimensions, including:
(i) sequence plausibility based on learned models,
(ii) structural metrics such as interface RMSD and pLDDT confidence,
(iii) physicochemical properties like charge complementarity and interface hydrophobicity.
This comprehensive analysis allows for more nuanced prioritization, identifying candidates with both strong binding affinity and structural stability.
\item \textbf{Integrated and User-Friendly Platform}:
The platform integrates multiple design and analysis tools into a streamlined workflow, eliminating the need for users to assemble components individually. With minimal configuration required, researchers can complete full design cycles effortlessly. Via the \href{https://paddlehelix.baidu.com/app/all/helixdesign-binder/forecast}{PaddleHelix} platform, users can easily launch binder design tasks and interact with predicted structures through an intuitive online interface.
\end{itemize}

\section{Method}
\label{sec:method}

\begin{figure}
\centering
\includegraphics[width=1.0\linewidth]{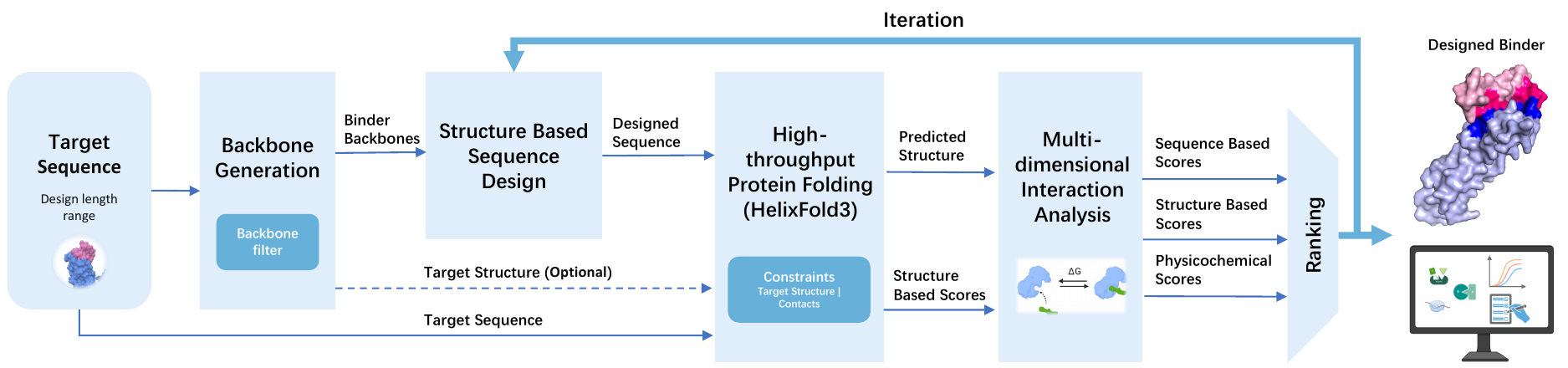}
\caption{HelixDesign-Binder for structure-based binder design. HelixDesign-Binder begins with a user-defined target sequence within a specified design length range. Candidate backbone conformations are generated and filtered by a backbone generation module. A structure-based sequence design module then proposes optimized amino acid sequences for the selected backbones. These candidate sequences undergo high-throughput structure prediction using HelixFold3, optionally guided by structural or residue-level contact constraints. The predicted models are subsequently evaluated through multi-dimensional interaction analysis (e.g., binding energy, interface features). Final binder candidates are prioritized via a ranking module that integrates structural and energetic metrics.}
\label{fig:framework} 
\end{figure}

\begin{figure}[htbp]
    \centering
    \includegraphics[width=\linewidth]{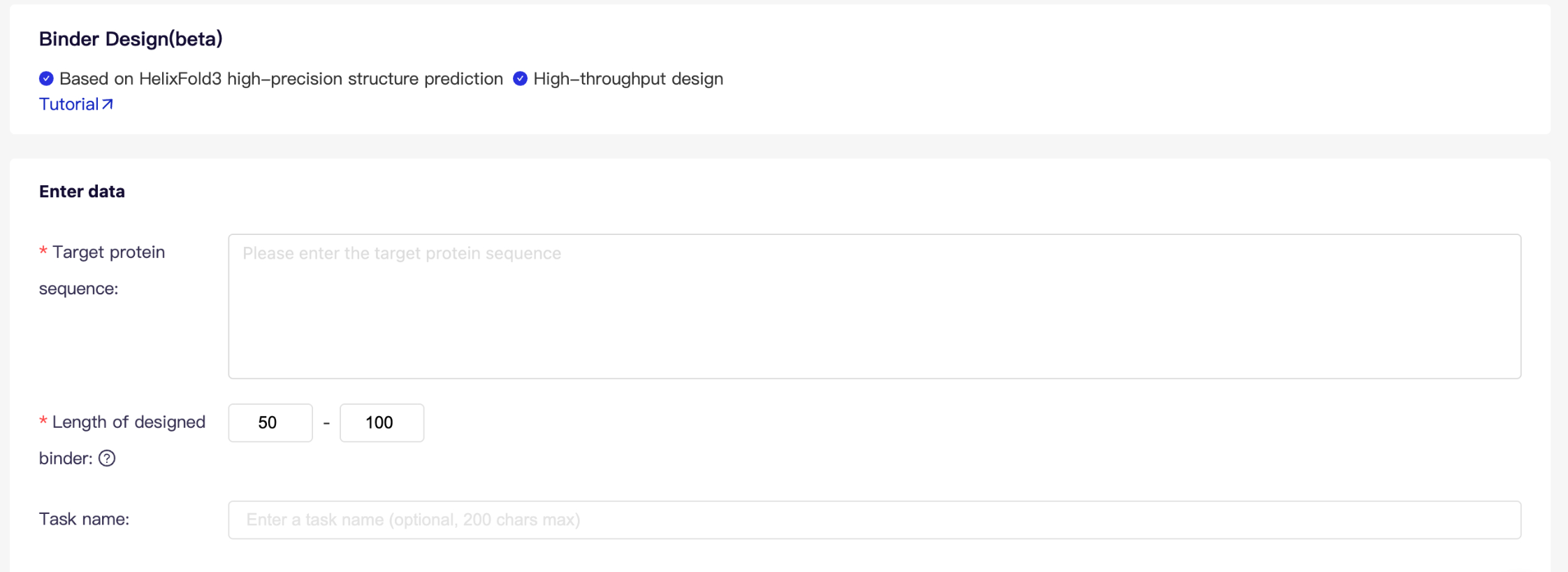}
    \caption{Example input interface of the HelixDesign-Binder Server.
    }
    \label{fig:input_server}
\end{figure}

\begin{figure}[htbp]
    \centering
        \includegraphics[width=\linewidth]{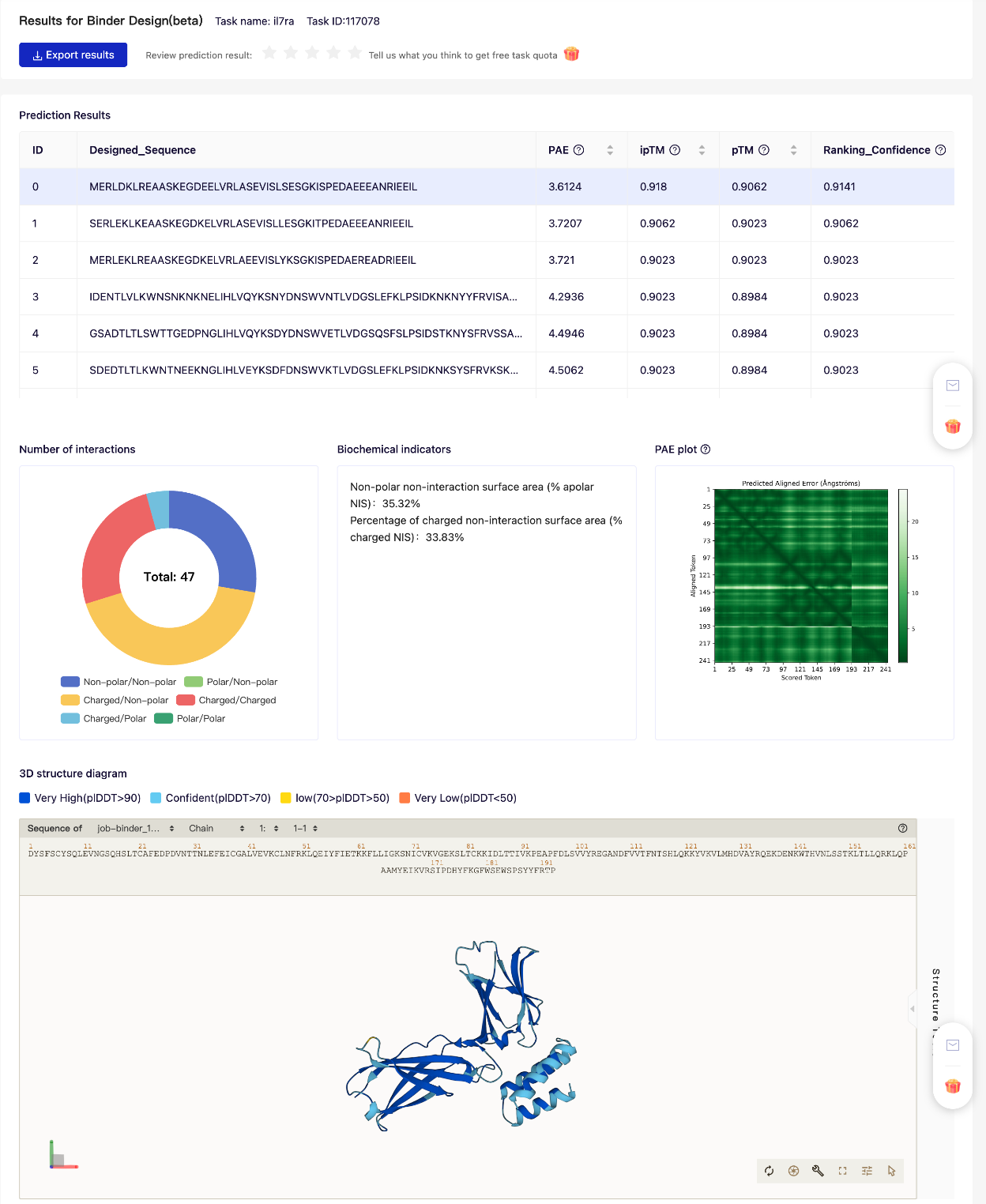}        
    \caption{Example output display page of the HelixDesign-Binder Server.
    }
    \label{fig:output_server}
\end{figure}

HelixDesign-Binder (Figure~\ref{fig:framework}) is a scalable production-grade platform designed for the efficient generation and evaluation of protein binder candidates targeting a specified sequence.
Existing design approaches \cite{watson2023novo, bennett2023improving, cao2022design} typically require the generation of thousands to millions of candidate sequences. Even methods employing iterative sampling strategies \cite{pacesa2024bindcraft} still necessitate hundreds to thousands of samples. These candidates must subsequently undergo structure prediction for downstream filtering, resulting in considerable computational overhead. To address this challenge, HelixDesign-Binder is optimized for high-throughput workflows in high-performance computing (HPC) environments, enabling the rapid screening of thousands of candidate binders within a single iteration.
It accommodates flexible design specifications, including both linear peptides and small proteins, making it well-suited for applications requiring high binding affinity and precise interface geometry.
As shown in Figure \ref{fig:input_server}, users can specify the desired sequence length range of the binder according to their design objectives. Based on these inputs, HelixDesign-Binder systematically constructs and ranks candidate sequences with strong potential for high-affinity recognition of the target, thereby enabling an efficient and focused search across a large and diverse sequence space.
Figure~\ref{fig:output_server} presents the design results generated by HelixDesign-Binder, along with the multi-dimensional interaction analysis output, which facilitates downstream interpretation and application by users.

The HelixDesign-Binder workflow comprises the following major components:
\begin{itemize}
	\item Backbone Generation: Initial structural backbones of the binder are generated based on the user-specified design length range and target sequence. A backbone filter module selects diverse, plausible scaffold candidates suitable for interface formation.
	\item Structure-Based Sequence Design: Given the selected backbones, this module employs inverse folding techniques to design amino acid sequences compatible with the backbone geometry and the target interface. 
	\item High-throughput Protein Folding (HelixFold3): Designed sequences are folded in complex with the target using HelixFold3, which supports the incorporation of external structural constraints. This step evaluates structural plausibility and predicts the full complex conformation.
	\item Multi-dimensional Interaction Analysis: Predicted complex structures are assessed across several dimensions, including sequence-based fitness, structure-based interface metrics, and physicochemical interaction scores. These scores are used to rank and prioritize candidate designs.
\end{itemize}

\subsection{Backbone Generation}
%Given a user-defined target sequence and the specified length range for the binder, this module generates a structurally diverse ensemble of candidate backbone conformations. These conformations provide the fixed structural frameworks upon which sequence design is subsequently performed. To ensure both structural feasibility and effective target engagement, a dedicated filtering step evaluates candidate backbones based on geometric plausibility, predicted interface quality, and overall conformational diversity. This process promotes broad sampling of the structural space while prioritizing backbones that are compatible with high-affinity binding.

In the backbone generation stage, we generate a diverse ensemble of candidate binder scaffolds based on the user-provided target and specified binder length range. Rather than constructing these backbones de novo, we leverage structural data from the Protein Data Bank (PDB) \cite{burley2017protein} by identifying complexes that are structurally similar to the input target. From these complexes, we extract interface fragments that meet the spatial and length constraints defined by the user, treating them as potential binder backbones. To ensure their structural quality and relevance, we evaluate these candidates across multiple dimensions—including sequence and geometric similarity to the target, structural plausibility as assessed by HelixFold3, and overall conformational diversity. This approach enables broad yet focused exploration of the structural landscape, providing reliable starting points for downstream sequence design.

\subsection{Structure-Based Sequence Design}
At this stage, binder amino acid sequences are generated based on the target proteins and binder backbones produced in the previous step. We utilize the ESM-IF1 inverse folding model \cite{hsu2022learning}, which generates sequences conditioned on backbone structures by estimating the probability distribution of residues at each position, informed by both local and global structural features. Because binder design involves at least two protein chains—the target and the binder—the original ESM-IF1 model, which supports sequence generation only for single chains, was modified in its inference procedure to accommodate multi-chain protein complexes.

To ensure broad structural coverage, sequences are generated approximately evenly across all selected backbones. The design module uses backbone constraints to optimize residue identities within each scaffold. For each designed sequence, a fitness score is computed from the inverse folding model outputs to quantify its compatibility with the corresponding backbone. These scores guide an initial filtering step that retains high-quality sequences while maintaining sequence diversity. For each target, at least 1,000 sequences passing this filter advance to the next stage, enabling thorough exploration of the sequence space. In future applications, users will have the flexibility to specify the number of sequences to be designed according to their experimental needs. Generating such a large number of candidate sequences is essential because binder design inherently involves a vast and complex sequence landscape; consistent with prior studies \cite{bennett2023improving, hsu2022learning, dauparas2022robust, hayes2025simulating}, producing thousands to millions of initial sequences before structural filtering.

\subsection{High-throughput Protein Folding (HelixFold3)}
Designed sequences are modeled in complex with the target using HelixFold3 \cite{liu2024technical}, our independently developed biomolecular structure prediction model. HelixFold3 achieves accuracy comparable to AlphaFold3 and surpasses it in certain specialized scenarios. It produces full atomic-resolution 3D structures of predicted complexes along with quantitative confidence metrics such as the interface predicted TM-score (ipTM), which facilitate reliable assessment of interface quality. In addition, HelixFold3 supports the incorporation of reference structures and interaction constraints, allowing the integration of prior knowledge to further enhance prediction accuracy and reliability.

To meet the computational demands of high-throughput binder screening, we leverage the high-performance computing (HPC) platform provided by Baidu AI Cloud. We have conducted extensive performance optimizations targeting critical stages such as multiple sequence alignment (MSA) search and structure prediction. By fully exploiting the parallel computing capabilities of the HPC platform and implementing optimization strategies tailored to the varying resource requirements and runtime characteristics of different modules, we achieve efficient utilization of computational resources and rapid task execution. These optimizations significantly improve overall throughput, enabling the rapid exploration and screening of large sequence spaces.

\subsection{Multi-dimensional Interaction Analysis}

To identify high-quality binder candidates, HelixDesign-Binder employs a comprehensive multi-dimensional evaluation framework comprising sequence-based, structure-based, and energy-based assessment methodologies.
Sequence-based evaluation utilizes protein language models trained on extensive natural sequence repositories to quantify the statistical likelihood and evolutionary fitness of each design. This assessment incorporates fitness scores derived from ESM-IF1\cite{hsu2022learning} calculations across the complete sequence length, providing estimates of biological plausibility based on learned evolutionary patterns.
Structure-based evaluation leverages HelixFold3-derived structural metrics to assess the physical feasibility and binding specificity of predicted protein-protein interfaces. Key metrics include the interface predicted TM-score (ipTM) for structural similarity assessment, inter-chain predicted aligned error (PAE) for confidence evaluation of inter-molecular interactions, and geometric contact maps for spatial relationship validation.
Physicochemical evaluation employs the computational tool PRODIGY\cite{xue2016prodigy} to estimate critical physicochemical properties governing binding interactions. These  scoring metrics encompass predicted binding affinity, percentage of apolar non-interface surface (NIS) residues, and comprehensive contact interface statistics including the total number of intermolecular contacts and the frequency of charged-apolar contact pairs.

Each candidate is filtered and ranked based on these sequence, structure, and physicochemical metrics. Final selection prioritizes designs that demonstrate strong binding potential while maintaining high structural fidelity to the intended interaction interface.
% To identify high-quality binder candidates, HelixDesign-Binder integrates a suite of complementary evaluation strategies. Sequence-based assessments leverage protein language models trained on large-scale natural sequence datasets to estimate the statistical plausibility and evolutionary fitness of each design. Structure-based evaluations utilize HelixFold3-derived metrics, including the interface predicted TM-score (ipTM), inter-chain predicted aligned error (PAE), and geometric contact maps, to assess the physical plausibility and specificity of the predicted interfaces. In addition, energy-based computational tools are employed to estimate key physicochemical properties, such as binding free energy and electrostatic complementarity at the interface.

%Each candidate is assigned a composite score that integrates these sequence, structure, and physicochemical metrics. Final selection prioritizes designs that demonstrate strong binding potential while maintaining high structural fidelity to the intended interaction interface. The pipeline further supports iterative refinement workflows, enabling repeated rounds of design and evaluation to progressively improve binder quality and convergence toward optimal candidates.

\section{Results}
\label{sec:results}
\begin{figure}[!t]
    \centering
    \begin{subfigure}[t]{0.48\linewidth}
        \centering
        \includegraphics[height=5.0cm]{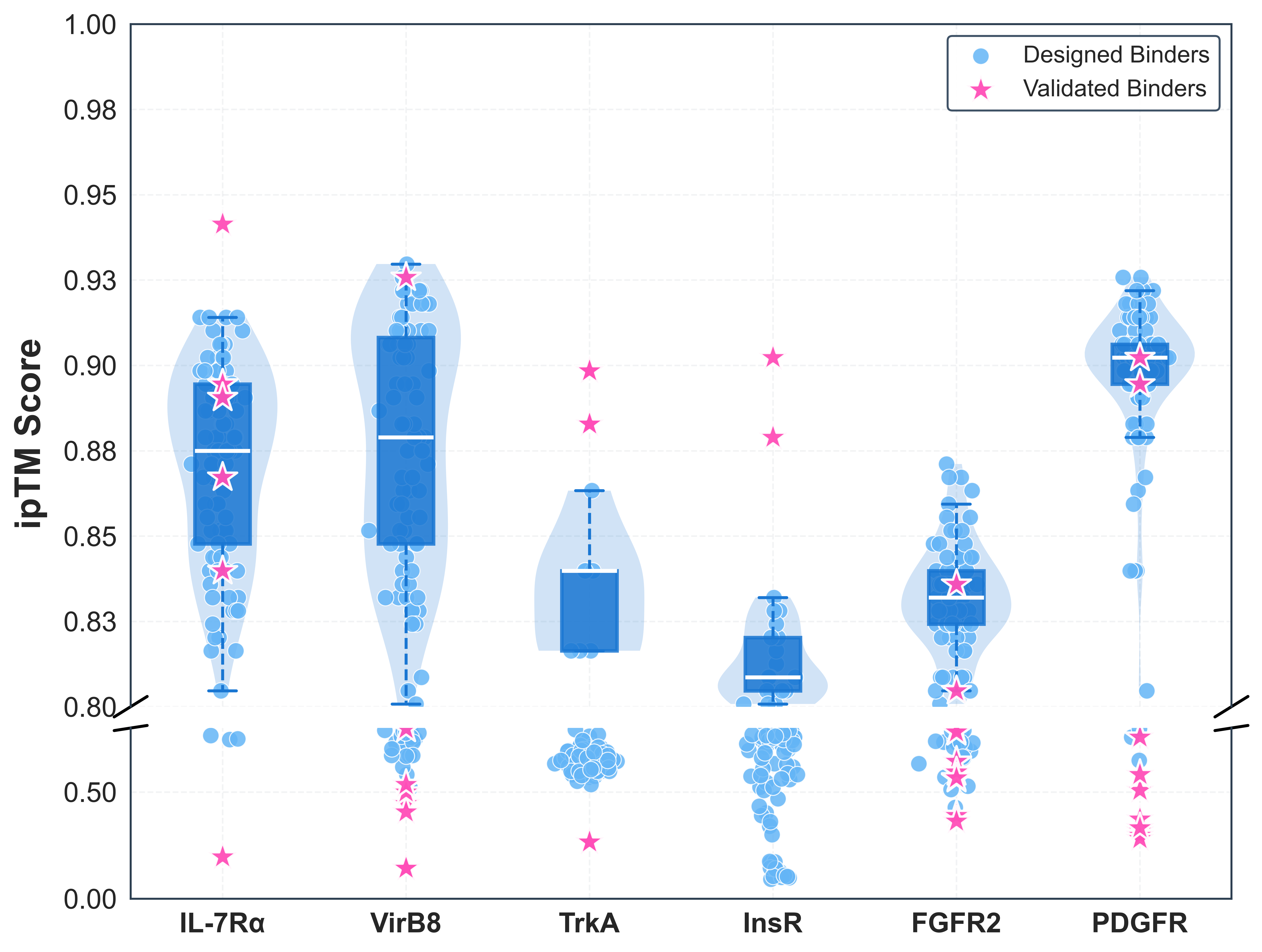}
        \caption{\hspace{1.2em}(a) Interface predicted TM-score (ipTM)}
        \label{fig:iptm}
    \end{subfigure}
    \hfill
    \begin{subfigure}[t]{0.48\linewidth}
        \centering
        \includegraphics[height=5.0cm]{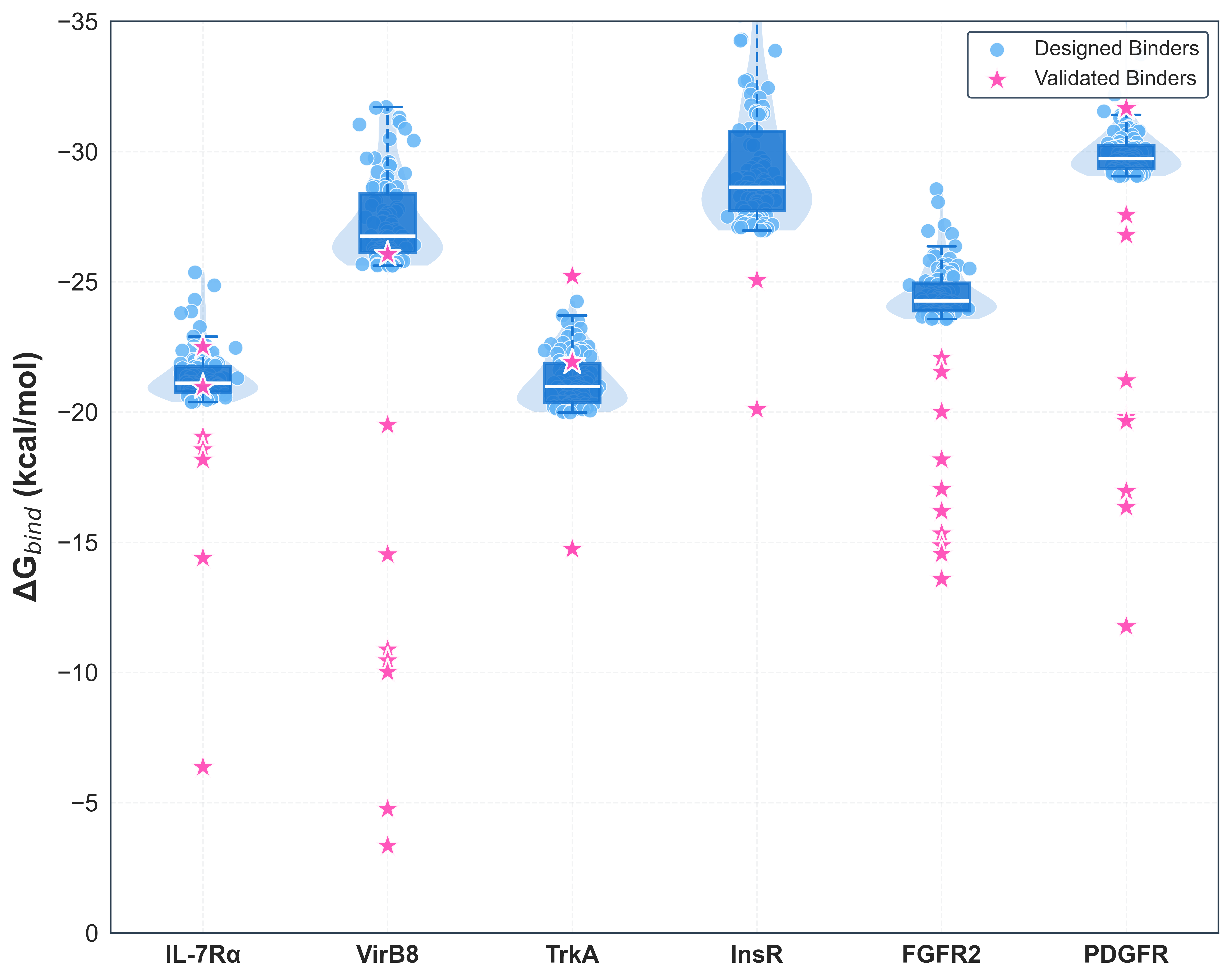}
        \caption{\hspace{1.2em}(b) Predicted binding free energy}
        \label{fig:dgfoldx}
    \end{subfigure}
    
    \vspace{0.5cm}

    \begin{subfigure}[t]{0.50\linewidth}
        \centering
        \includegraphics[height=5.21cm]{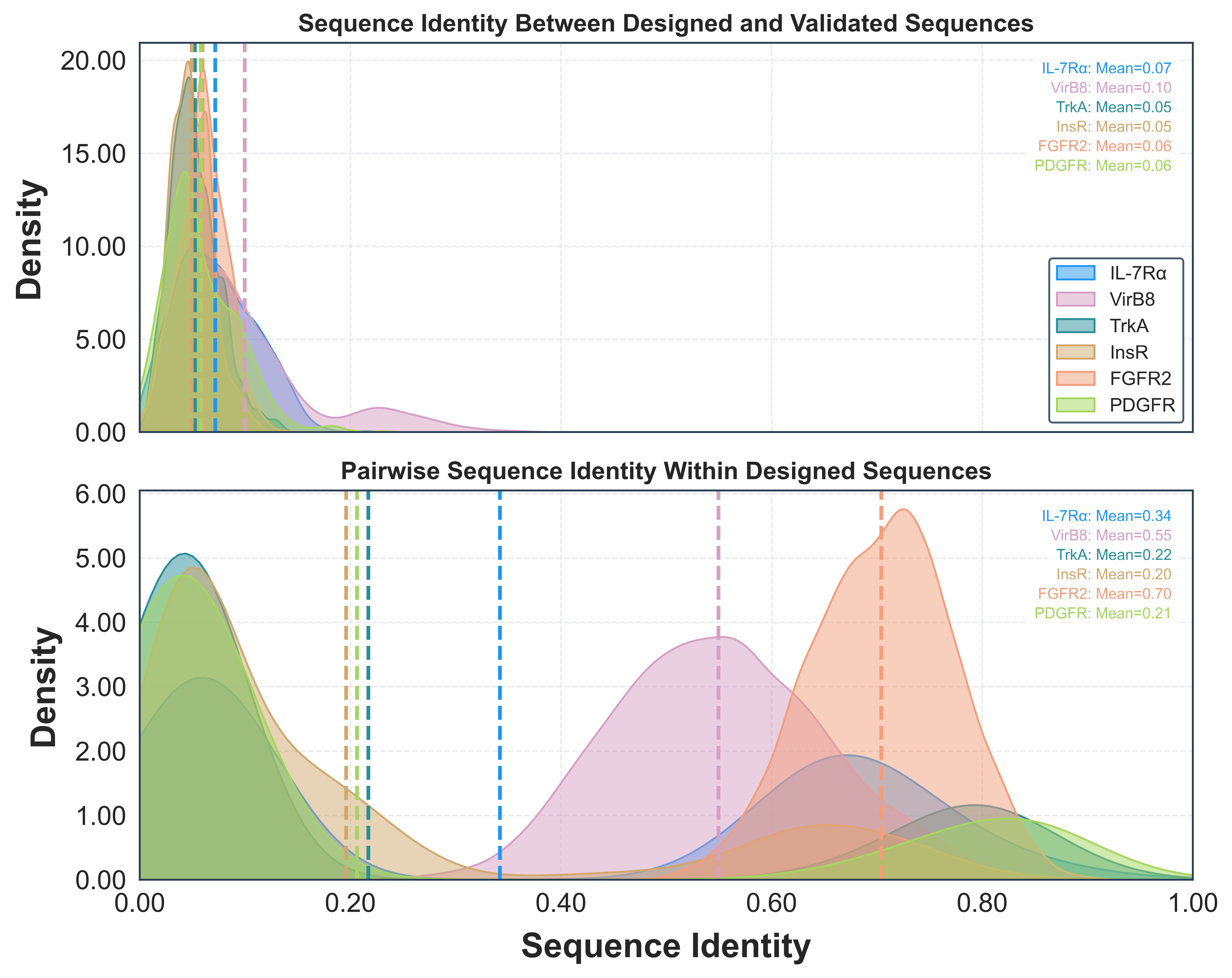}
        \caption{\hspace{1.2em}(c) Sequence identity analysis of designed binders}
        \label{fig:pairwise_identity}
    \end{subfigure}
    \hfill
    \begin{subfigure}[t]{0.48\linewidth}
        \centering
        \includegraphics[height=5.05cm]{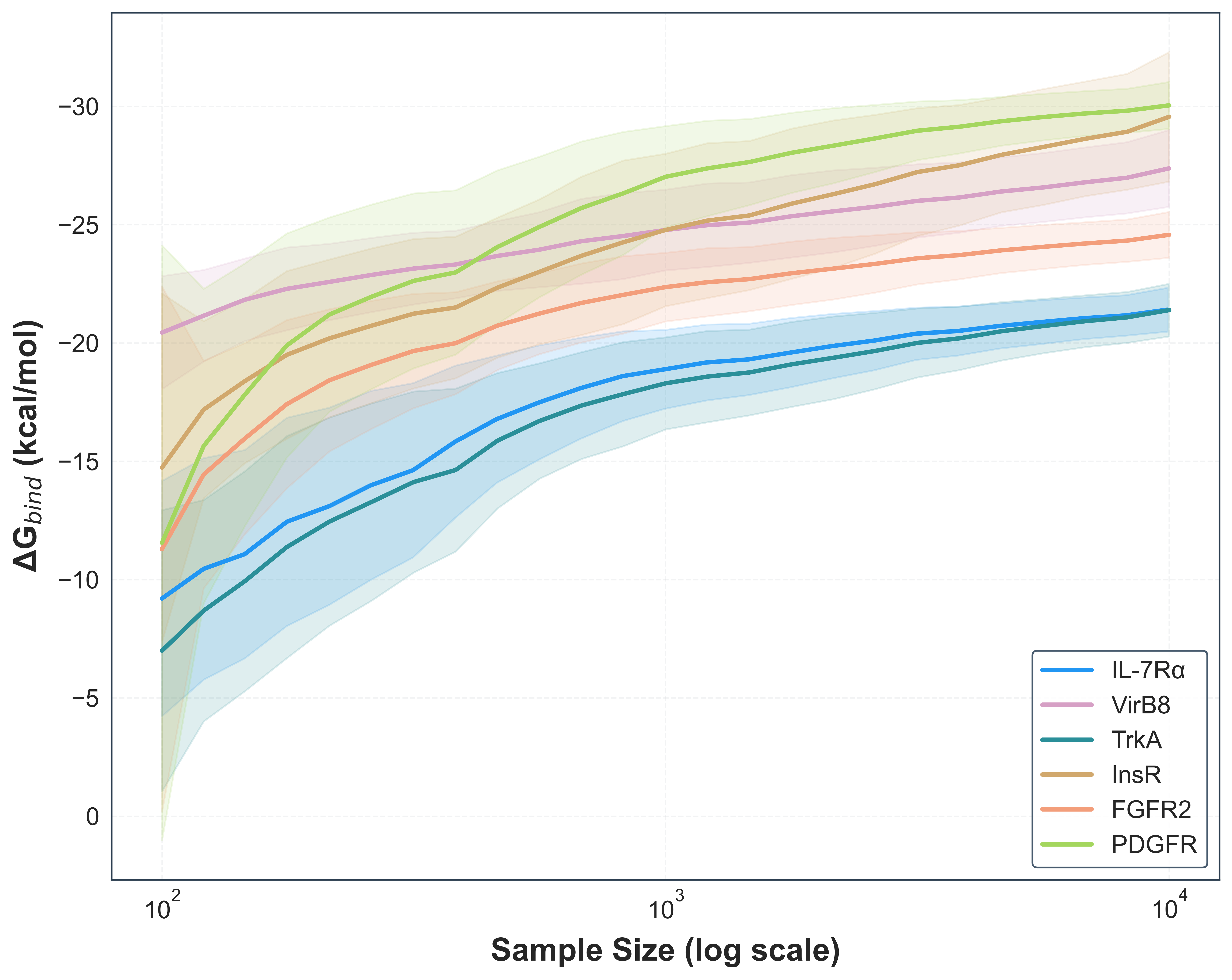} 
        \caption{\hspace{1.2em}(d) Improvement in binding free energy as a function of sampling size}
        \label{fig:other}
    \end{subfigure}

    \vspace{0.5cm}

    \begin{subfigure}[t]{0.9\linewidth}
    \includegraphics[width=\linewidth]{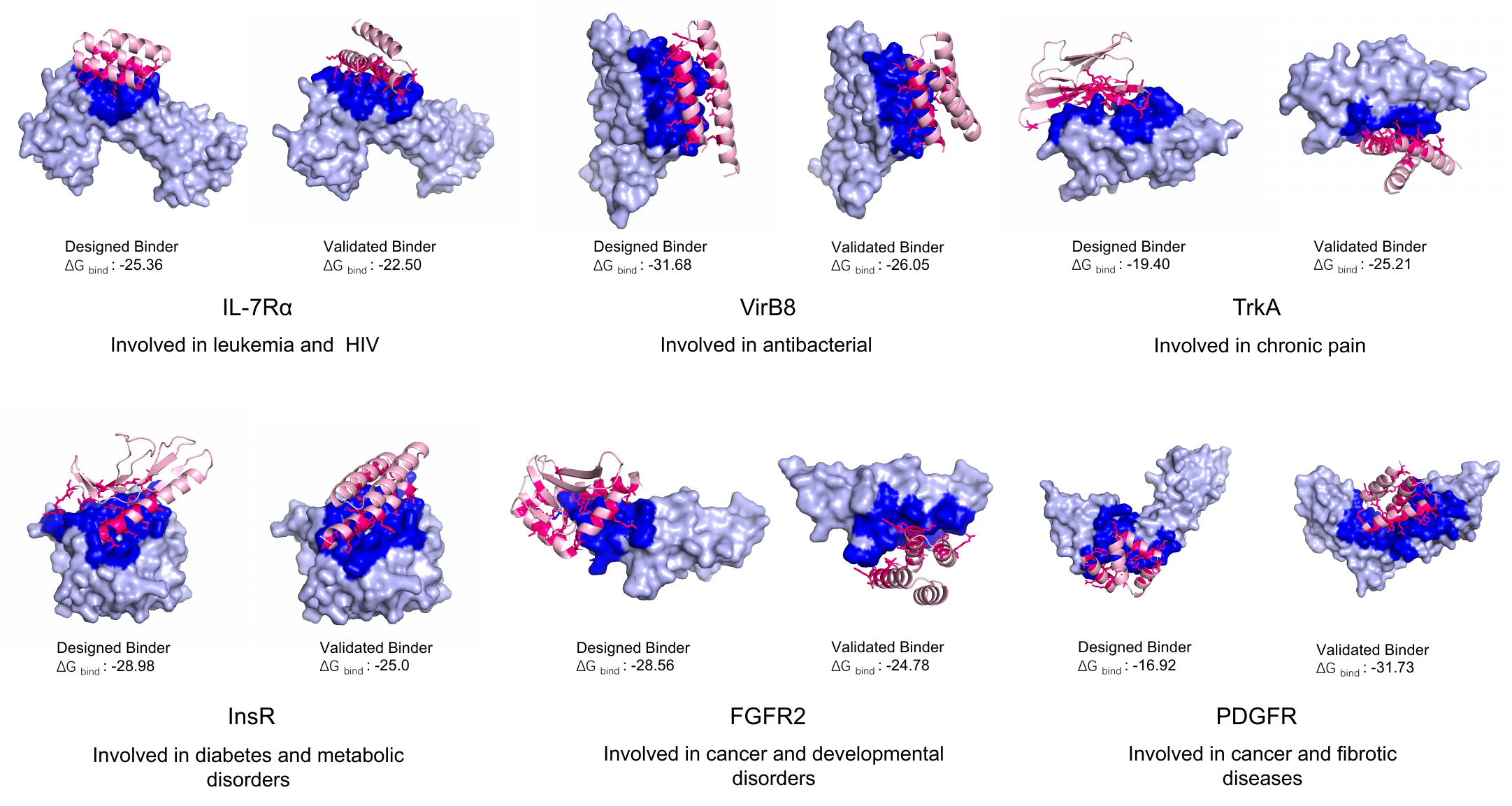}
    \caption{\hspace{1.2em}(e) Structural visualization of designed and validated binders}
    \label{fig:cases}
    \end{subfigure}
    
    \caption{Results of HelixDesign-Binder on six protein targets.}
    \label{fig:comparison}
\end{figure}

\begin{figure}[htbp]
    \centering

    % 第一行
    \begin{subfigure}[b]{0.33\linewidth}
        \includegraphics[width=\linewidth]{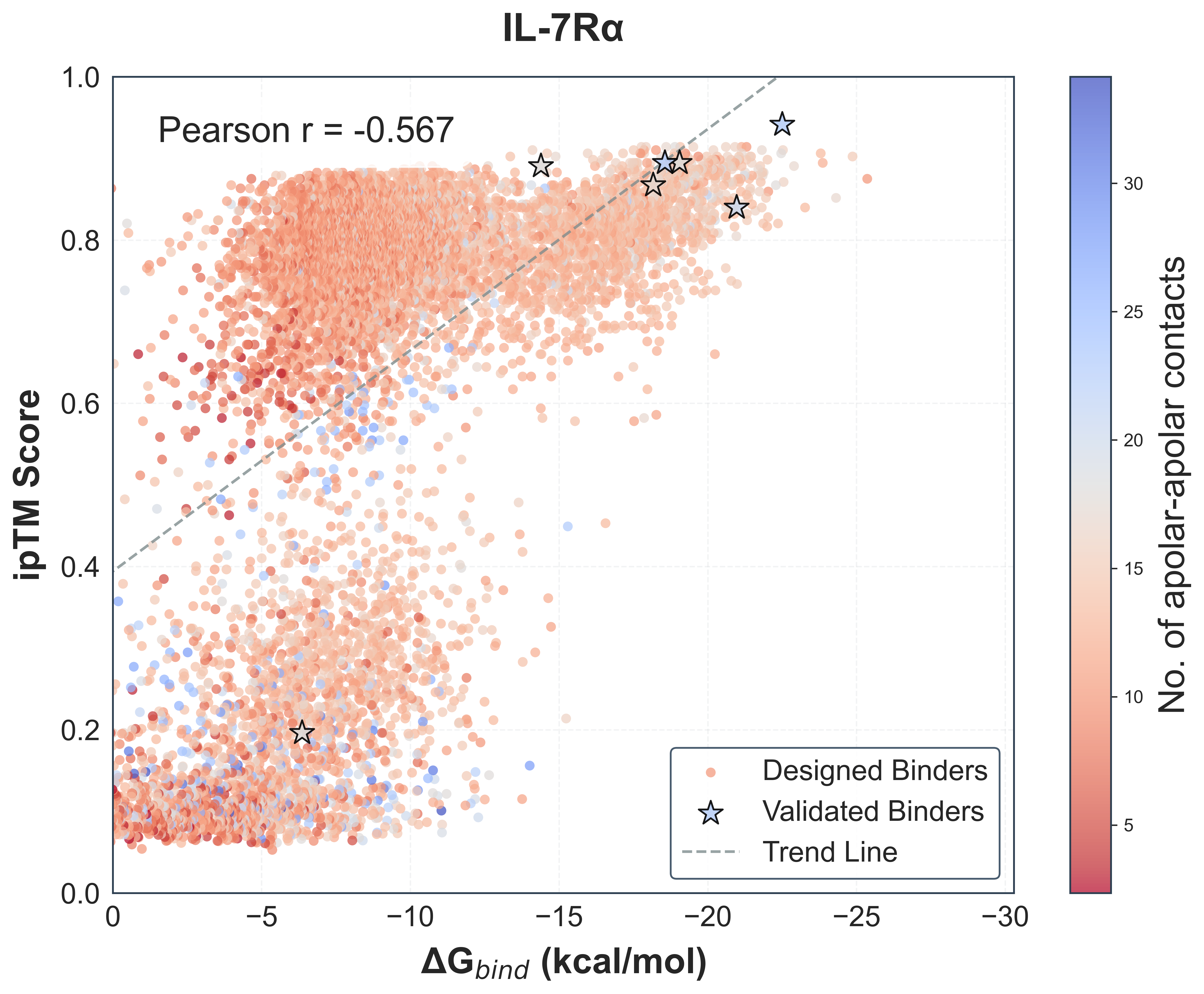}
        \caption{(a)}
        \label{fig:sub1}
    \end{subfigure}
    \hfill
    \begin{subfigure}[b]{0.33\linewidth}
        \includegraphics[width=\linewidth]{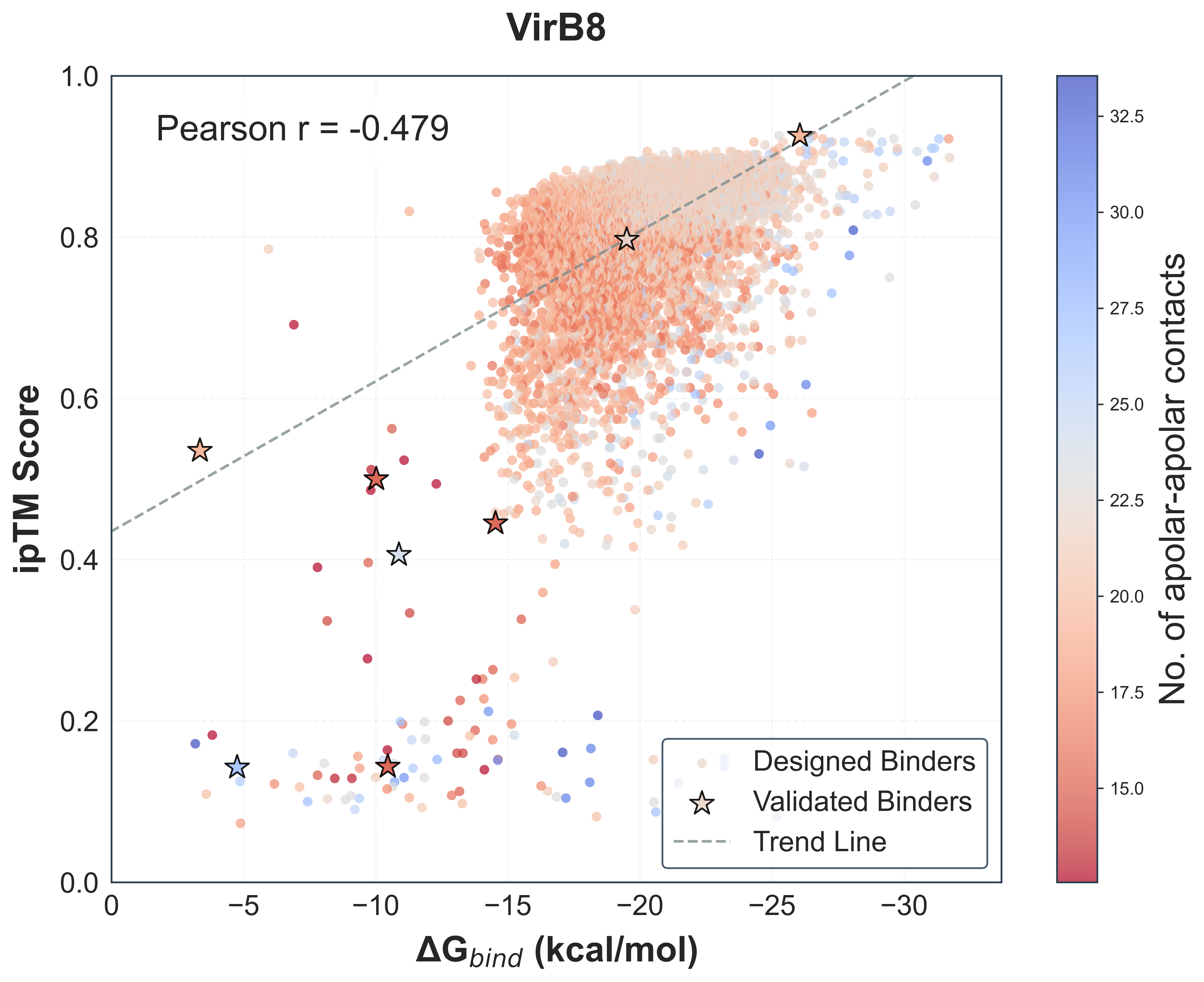}
        \caption{(b)}
        \label{fig:sub2}
    \end{subfigure}
    \hfill
    \begin{subfigure}[b]{0.33\linewidth}
        \includegraphics[width=\linewidth]{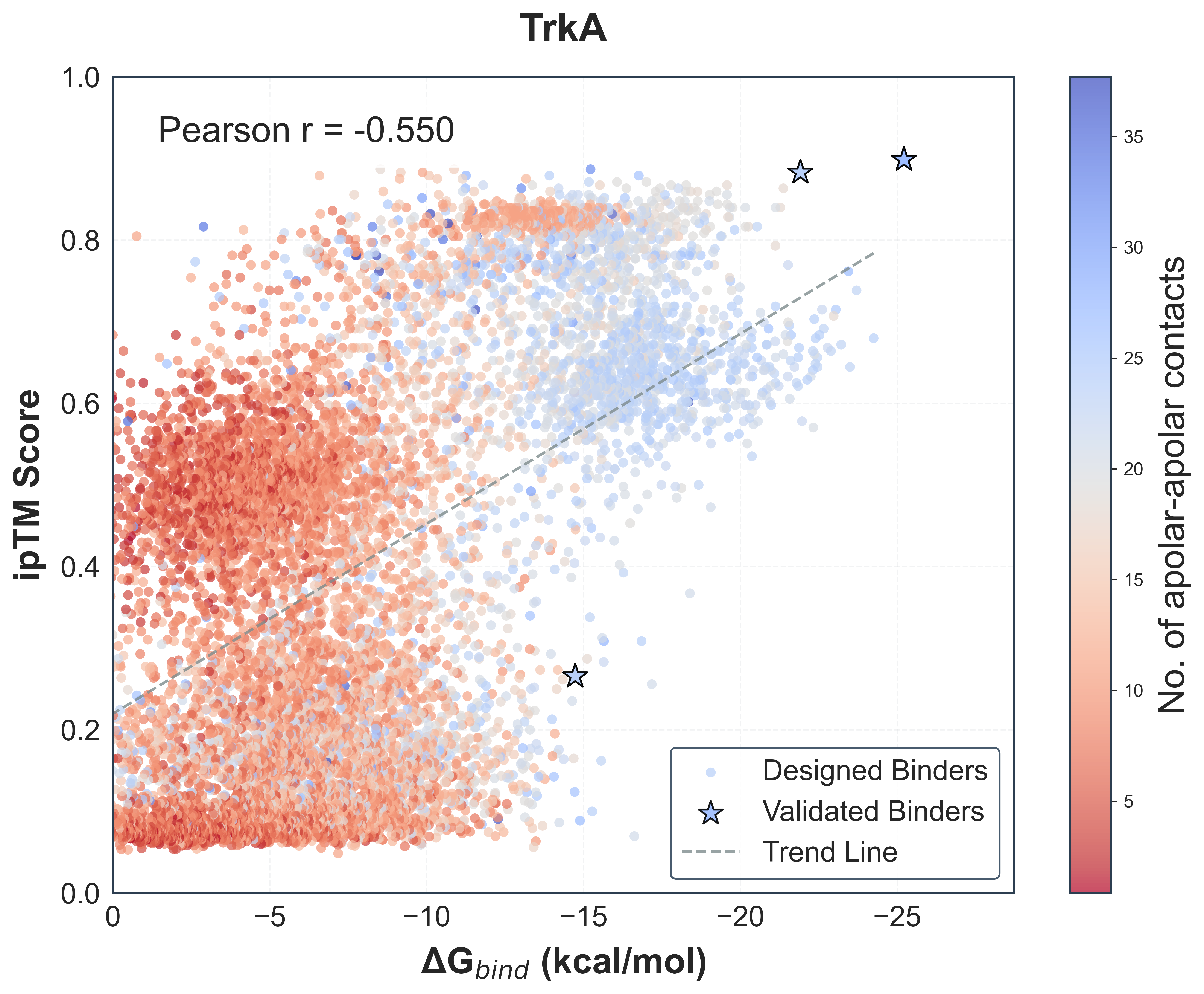}
        \caption{(c)}
        \label{fig:sub3}
    \end{subfigure}

    \vspace{1em}  % 行间距
    % 第二行
    \begin{subfigure}[b]{0.33\linewidth}
        \includegraphics[width=\linewidth]{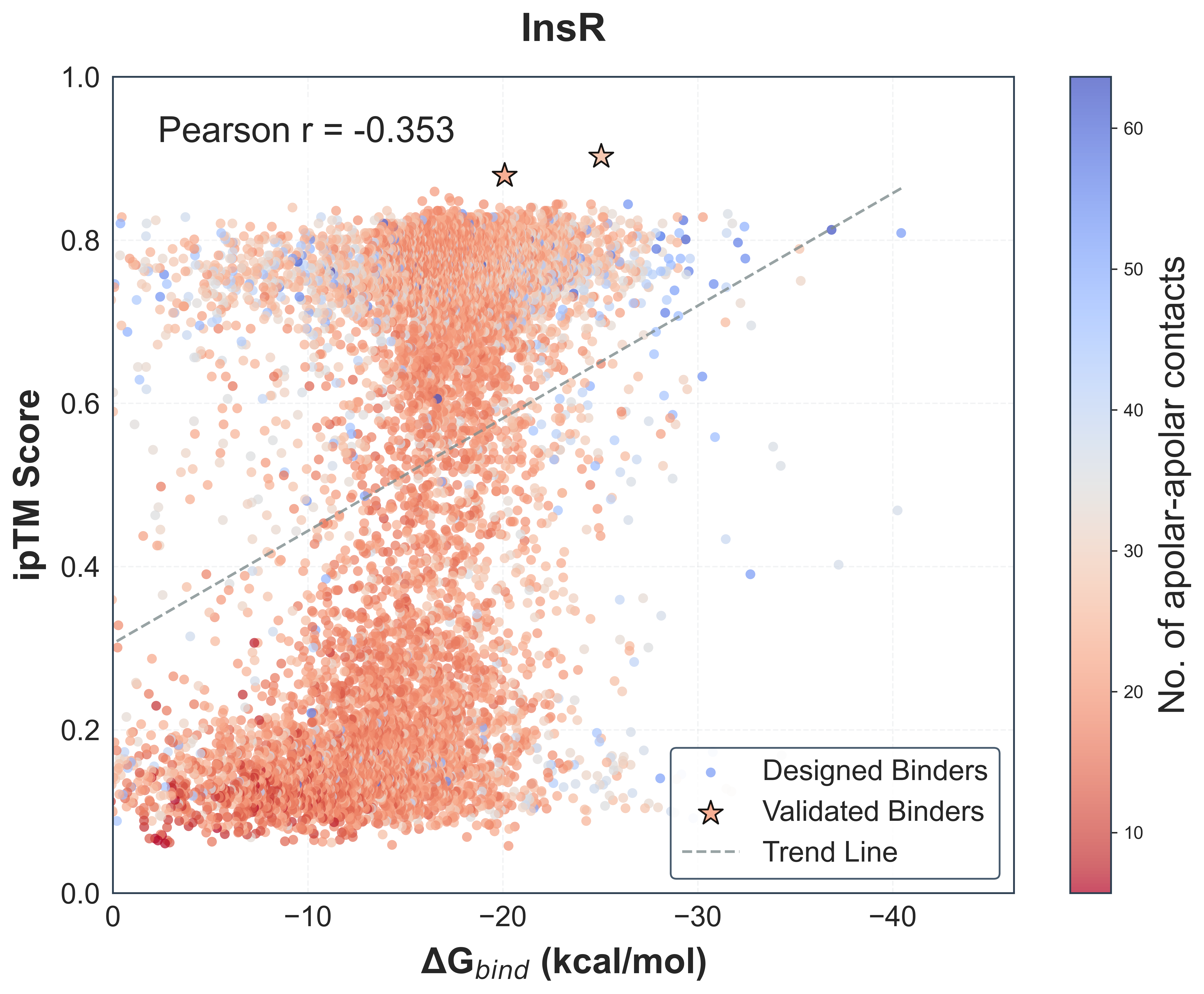}
        \caption{(d)}
        \label{fig:sub4}
    \end{subfigure}
    \hfill
    \begin{subfigure}[b]{0.33\linewidth}
        \includegraphics[width=\linewidth]{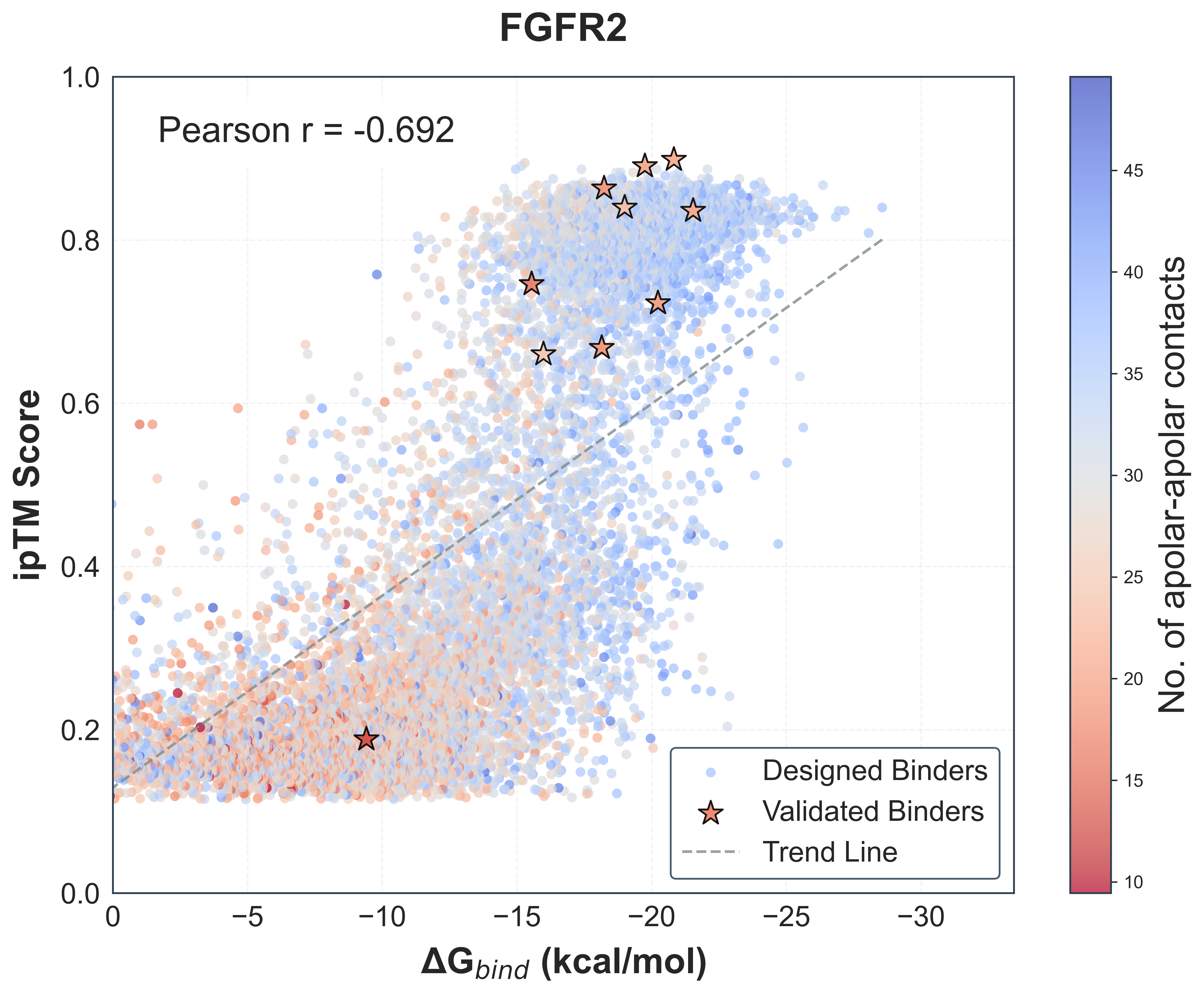}
        \caption{(e)}
        \label{fig:sub5}
    \end{subfigure}
    \hfill
    \begin{subfigure}[b]{0.33\linewidth}
        \includegraphics[width=\linewidth]{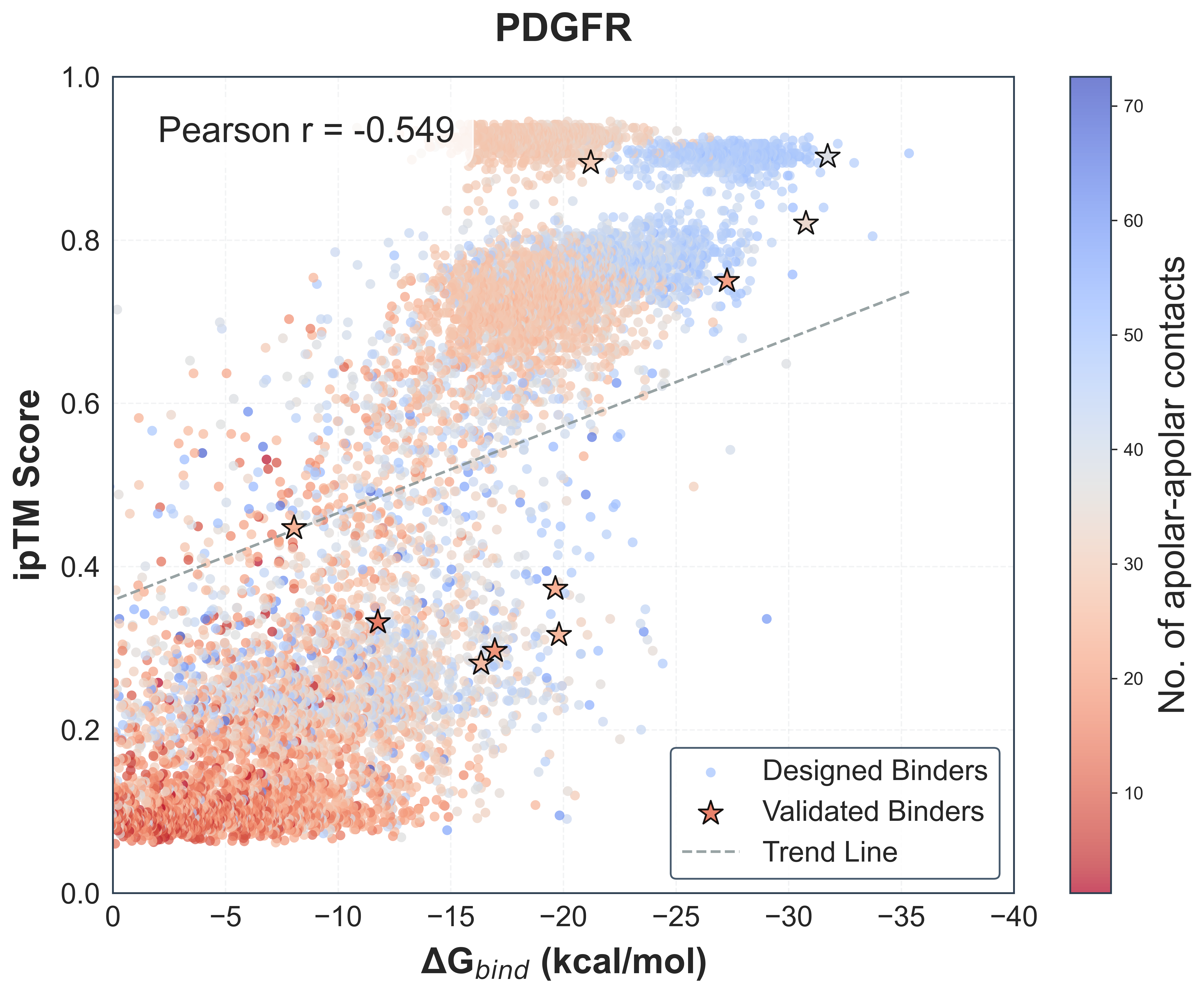}
        \caption{(f)}
        \label{fig:sub6}
    \end{subfigure}

    \caption{Correlation between predicted binding free energies (kcal/mol) and ipTM score for six protein targets: (a) IL1RA, (b) VirB8, (c) TrkA, (d) InsR, (e) FGFR2, and (f) PDGFR. Each point represents a binder, colored by the number of apolar-apolar contacts. Dots indicate designed binders, the stars represent the experimentally validated binders. Pearson correlation coefficients reveal varying relationships across targets, with values ranging from -0.353 (InsR) to -0.692 (FGFR2).}
    \label{fig:scatter}
\end{figure}
% \end{appendices}

\subsection{Binder Design}

To rigorously evaluate the effectiveness of HelixDesign-Binder, we selected six previously characterized protein targets from \cite{cao2022design} as a validation benchmark. Our evaluation focused on two complementary metrics: the interface predicted TM-score (ipTM) and FoldX-predicted binding free energy. These metrics have been widely used and validated in prior studies as proxies for binding efficacy \cite{wee2024evaluation, patel2021implementing, gonzalez2020assessment, sirin2016ab}. Specifically, ipTM reflects the model's structural confidence at the predicted binding interface (with higher values indicating greater reliability), while FoldX estimates the binding free energy of the protein complex (with lower values indicating stronger predicted binding affinity). Together, these two measures provide a robust framework for assessing both structural plausibility and energetic favorability of designed binders.

Across the six targets: Interleukin-7 Receptor-\ensuremath{\alpha} (IL-7R\ensuremath{\alpha})\cite{mcelroy2009structural}, Virulence factor B8 (VirB8)\cite{gillespie2015structural}, Tropomyosin Receptor Kinase A (TrkA)\cite{wiesmann1999crystal}, Insulin Receptor (InsR)\cite{croll2016higher}, Fibroblast Growth Factor Receptor 2 (FGFR2)\cite{plotnikov2000crystal},  Platelet Derived Growth Factor Receptor (PDGFR)\cite{hye2010structures}, our computational approach demonstrated clear performance advantages. As positive controls, we used experimentally validated binders with confirmed binding affinity from the original study, hereafter referred to as validated binders. For targets with more than ten validated binders, we randomly selected a subset of ten for comparison.

As shown in Figure~\ref{fig:comparison}(a) and (b), the designed binders consistently achieved ipTM scores above 0.8, indicating that the structure prediction model considered these binder–target complexes to be highly reliable. Previous studies have demonstrated that ipTM scores are positively correlated with binding capability, supporting the relevance of this metric in evaluating binder quality \cite{bryant2022improved,bennett2023improving,kim2024enhanced}. For four targets (IL-7R\ensuremath{\alpha}, VirB8, FGFR2, and PDGFR), the designed binders achieved ipTM scores comparable to or higher than those of validated binders. For the other two targets (TrkA and INSULNR), the designed binders exhibited slightly lower ipTM scores than their validated counterparts. From a thermodynamic perspective, FoldX analysis indicated that the designed binders exhibited more favorable (i.e., lower) predicted binding free energies than the validated binders across all targets except TRKA. Notably, binders designed for VIRB8 demonstrated particularly favorable energetics, with mean Gibbs free energy values approaching –27.3 kcal/mol—substantially lower than those of the corresponding validated binders (-12.4 kcal/mol).

While ipTM and FoldX-predicted binding energy correlate with binding activity, they are not definitive predictors, some validated binders score poorly on one or both axes. Nonetheless, binders with favorable values for both metrics show a markedly higher likelihood of being active, making them valuable for candidate prioritization.

%This dual achievement, maintaining structural confidence while enhancing binding energetics, was consistently observed across most of target proteins, with designed binders maintaining mean interaction energy values between -15 and -22 kcal/mol, indicating robust thermodynamic profiles. Moreover, statistical distribution analysis revealed that the designed binders showed less variability in both metrics, whereas validated binders exhibited broader distributions and more outliers, especially evident in Figure~\ref{fig:comparison} (a, b). We attribute this consistency to the ranking strategy employed during binder selection, which integrates multi-dimensional interaction analyses. This suggests that a larger fraction of our designed binder ensemble may not only adopt the correct binding orientation but also possess experimentally measurable affinity improvements.

\subsection{Design Space Exploration and Functional Performance}
% Our HelixDesign-Binder framework exhibits a strong capacity for extensive sequence space exploration, which plays a key role in identifying high-quality binders with favorable biophysical properties. 
% As shown in the pairwise sequence identity distributions (Figure~\ref{fig:comparison}(c)), the bottom panel demonstrates that the designed binders span a wide range of sequence similarity, indicating substantial intra-group diversity. Meanwhile, the top panel reveals that the designed sequences exhibit very low similarity to experimentally validated binders from previous studies (sequence identity  $\leq 0.1$ )\cite{cao2022design}. Together, these results suggest that the framework systematically explores structurally viable yet sequence-diverse solutions.
% Distinct target-specific patterns are evident: IL-7R\ensuremath{\alpha}, VirB8, and FGFR2 designs tend to cluster more tightly, whereas TrkA, InsR, and PDGFR designs display broader identity distributions. These differences suggest that the search strategy adapts dynamically to each target’s underlying structural and functional landscape.

The HelixDesign-Binder workflow demonstrates strong capabilities in generating novel sequences while maintaining diversity within each design set, two essential factors for effective binder discovery. In terms of novelty, the top panel of Figure~\ref{fig:comparison}(c) shows that the designed binders exhibit extremely low sequence identity ($\leq0.1$) to experimentally validated binders from prior studies \cite{cao2022design}, indicating that the workflow actively explores unexplored regions of the sequence space rather than replicating known solutions. In terms of diversity, the bottom panel reveals broad pairwise sequence identity distributions within each target-specific design set, suggesting substantial intra-group variability. For all targets, a significant fraction of binder pairs share less than 20\% sequence identity, reflecting the framework’s capacity to generate structurally viable yet sequence-diverse candidates.

We visualized representative structures from the top-ranking region alongside experimentally validated binders and reported their predicted binding free energies (Figure~\ref{fig:cases}). The designed binders exhibit diverse secondary structure topologies—including alpha-helical, beta-sheet, and mixed conformations—particularly for targets such as TrkA and FGFR2, highlighting structural as well as sequence-level diversity. For three targets where the binding sites of designed and validated binders are spatially aligned, the predicted binding energies of the designed binders outperform those of the validated counterparts, suggesting that HelixDesign-Binder can generate candidates with potentially improved binding properties.

\subsection{Computational Scale Enables Binder Optimization}
% This extensive sequence exploration correlates with a consistent scaling relationship between sample size and binding performance (Figure~\ref{fig:comparison} (d)). As the number of sampled sequences increases logarithmically, the average predicted binding free energy improves across all six protein targets, revealing a clear trend that broader sampling enhances design quality. Moreover, we observed a concurrent increase in structural confidence: the mean ipTM scores of the top 100 predicted binders also rise with larger sampling sizes, reinforcing that deeper exploration not only improves binding energetics but also yields more reliable and interface-consistent structures. Together, these results confirm the presence of a quantitative scaling law in our pipeline, where increased sampling leads to more optimized designs both thermodynamically and structurally.

% However, this trend also highlights a key limitation of current design models\cite{bennett2023improving, hsu2022learning, dauparas2022robust,hayes2025simulating}: achieving high-quality results often requires massive sampling, which may not be computationally feasible for most practitioners. To address this challenge, our HelixDesign-Binder pipeline leverages high-performance computing (HPC) resources to support large-scale sampling, thereby enabling comprehensive exploration of sequence space that would otherwise be inaccessible in typical research settings.

Most successful protein design studies to date rely on massive sampling strategies, often generating thousands to millions of candidate sequences before applying downstream filtering \cite{bennett2023improving, hsu2022learning, dauparas2022robust, hayes2025simulating}. However, this approach poses a significant practical barrier: many prospective users of these design tools lack access to such computational scale and can only afford to generate a small number of designs, often resulting in suboptimal performance. To address this limitation, the HelixDesign-Binder is built to fully leverage high-performance computing (HPC) resources, enabling efficient large-scale sampling across diverse protein targets. This allows for comprehensive exploration of sequence space, improving the chances of identifying candidates with strong biophysical and structural properties even for challenging targets.

This extensive sampling is not only computationally feasible within our framework—it also yields clear performance benefits. As shown in Figure~\ref{fig:comparison}(d), we observe a consistent scaling relationship between the number of sampled sequences and predicted binding quality. Specifically, as sample size increases logarithmically, the average predicted binding free energy improves across all six targets, indicating enhanced thermodynamic favorability. At the same time, the mean ipTM scores of the top 100 binders also rise, reflecting greater structural confidence and interface consistency. Together, these results reveal a quantitative scaling law: larger sampling directly contributes to the discovery of more optimized binders, both in terms of binding energetics and structural reliability.

\subsection{Multi-dimensional Interaction Analysis}

To comprehensively evaluate binder quality, we adopt a multi-dimensional assessment framework, as different metrics capture distinct yet complementary aspects of binding performance. Specifically, we selected: (i) the interface predicted TM-score (ipTM) to assess structural confidence at the binding interface; (ii) predicted binding free energy calculated using FoldX\footnote{Note that the public online version of HelixDesign-Binder uses PRODIGY for binding free energy estimation due to its higher efficiency in web deployment}; and (iii) the number of hydrophobic contacts to quantify interface compactness and physicochemical complementarity \cite{klebe2025protein}. Together, these metrics reflect geometric plausibility, energetic favorability, and chemical compatibility, providing a comprehensive characterization of designed binders.

Across six protein targets, we observed a strong inverse correlation between ipTM and predicted binding free energy (Pearson r ranging from –0.35 to –0.69), indicating that designs with higher structural confidence at the interface tend to exhibit more favorable energetics. The imperfect correlation highlights their complementarity—ipTM emphasizes geometric consistency, while the predicted binding free energy evaluates atomistic physical interactions. Hydrophobic contacts further refine this assessment by identifying well-packed interfaces that may not be fully captured by geometry or energy alone. In the two-dimensional space defined by ipTM and binding free energy, designed binders are broadly distributed, with a notable subset enriched in hydrophobic contacts (indicated in blue) clustering in the upper-right quadrant, where both structural and energetic metrics are optimized. This region overlaps with experimentally validated binders (e.g., FGFR2 and IL-7R\ensuremath{\alpha}), suggesting that HelixDesign-Binder reliably identifies designs with favorable and balanced interaction profiles.

\section{Conclusion}
While the fundamental workflow for protein design—encompassing backbone generation, sequence design, and filtering—has become well-established, its practical application remains resource-intensive and complex for many users due to the multitude of tools, parameters, and high computational requirements involved. To lower these barriers and broaden accessibility, we developed HelixDesign-Binder, a scalable production-grade platform for protein binder design that features an intuitive, user-friendly visual interface. The platform’s effectiveness has been preliminarily validated across six diverse protein targets.

We invite researchers from both academia and industry to explore and evaluate HelixDesign-Binder via \href{https://paddlehelix.baidu.com/app/all/helixdesign-binder/forecast}{PaddleHelix platform}. We encourage feedback to identify challenges and guide ongoing improvements. Furthermore, we welcome collaboration opportunities to collectively advance the field of protein binder design. For inquiries and partnership discussions, please contact us at \href{mailto:baidubio_cooperate@baidu.com}{baidubio\_cooperate@baidu.com}.

% \section*{Acknowledgments}
% This was was supported in part by......

\clearpage

%Bibliography
\bibliographystyle{unsrt}  
\bibliography{references}

\clearpage
% \begin{appendices}

\end{document}